\documentclass{article}
\usepackage{spconf,graphicx,amsmath}
\usepackage[T1]{fontenc}
\usepackage[utf8]{inputenc}
\usepackage{subcaption}
\usepackage{booktabs}
\usepackage{enumitem}
\setlist{nosep,leftmargin=14pt}
\usepackage{amssymb}
\usepackage{tikz}
\usetikzlibrary{arrows.meta,positioning,shapes.geometric,fit,calc}
\definecolor{cbct}{HTML}{6EA8FE}
\definecolor{ios}{HTML}{FCC419}
\definecolor{proxy}{HTML}{8CE99A}
\definecolor{refine}{HTML}{FFA94D}
\tikzset{
  block/.style={rectangle,rounded corners,draw=black,thick,align=center,
                minimum height=1.05cm,minimum width=3.2cm,fill=white},
  small/.style={font=\scriptsize},
  line/.style={-Latex,thick},
  groupbox/.style={draw=black!50,rounded corners,inner sep=6pt,dashed}
}

\title{High-Fidelity 3D Tooth Reconstruction by Fusing Intraoral Scans and CBCT Data via Deep Implicit Representations}

\name{Yi Zhu$^{1, 2}$ \qquad Razmig Kéchichian$^{1}$ \qquad Raphaël Richert$^{3}$ \qquad
      Ikehata Satoshi$^{2}$ \qquad Sébastien Valette$^{1}$}

\address{$^{1}$ INSA-Lyon, UCBL, CNRS, Inserm, CREATIS UMR 5220, U1294, F-69621, Lyon, France \\
         $^{2}$ National Institute of Informatics (NII), Tokyo, Japan \\
         $^{3}$ Hospices Civils de Lyon, PAM Odontologie, Lyon, France
        }

\begin{document}

\maketitle
\begin{abstract}
High-fidelity 3D tooth models are essential for digital dentistry capturing both the detailed crown and the complete root. Clinical imaging modalities have limitations: Cone-Beam Computed Tomography (CBCT) captures the root but has a noisy, low-resolution crown, while Intraoral Scanners (IOS) provide a high-fidelity crown but no root information. A naive fusion of these sources results in seams and artifacts. We propose a novel, fully-automated workflow that fuses CBCT and IOS data using a deep implicit representation. Our method first segments and robustly registers tooth instances, then creates a hybrid proxy mesh combining the IOS crown and the CBCT root. The core of our approach is to use this noisy proxy to guide a class-specific DeepSDF network. This optimization process projects the input onto a learned manifold of ideal tooth shapes, generating a seamless, watertight, and anatomically coherent model. Qualitative and quantitative evaluations show that our method preserves both the high-fidelity crown from IOS and the patient-specific root morphology from CBCT, overcoming the limitations of each modality and naive stitching.
\end{abstract}


\section{Introduction}
\label{sec:intro}

The advent of digital dentistry has revolutionized clinical workflows in orthodontics, implantology, and prosthodontics. Central to this paradigm shift is the availability of accurate, 3D digital models of a patient's dentition. An ideal tooth model must capture two critical elements: \textbf{(1)} the high-fidelity surface geometry of the crown for designing restorations and appliances, and \textbf{(2)} the complete and precise morphology of the root for surgical planning, biomechanical simulation, and orthodontic treatment planning \cite{Scarfe2008CBCTHowWorks,Mangano2019IOSAccuracy,Alkadi2023IOSAccuracyReview}.

\begin{figure}[t]
    \centering
    \begin{subfigure}[b]{0.26\linewidth}
        \centering
        \includegraphics[width=\textwidth]{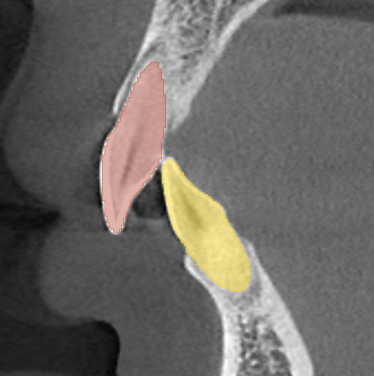}
        \caption{}
        \label{fig:cbct}
    \end{subfigure}
    \hfill
    \begin{subfigure}[b]{0.3\linewidth}
        \centering
        \includegraphics[width=\textwidth]{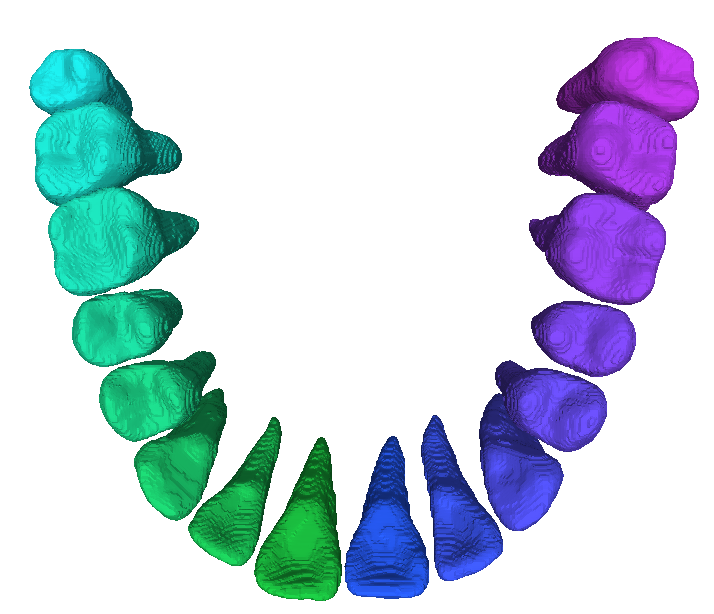}
        \caption{}
        \label{fig:cbct_mesh}
    \end{subfigure}
    \hfill
    \begin{subfigure}[b]{0.3\linewidth}
        \centering
        \includegraphics[width=\textwidth]{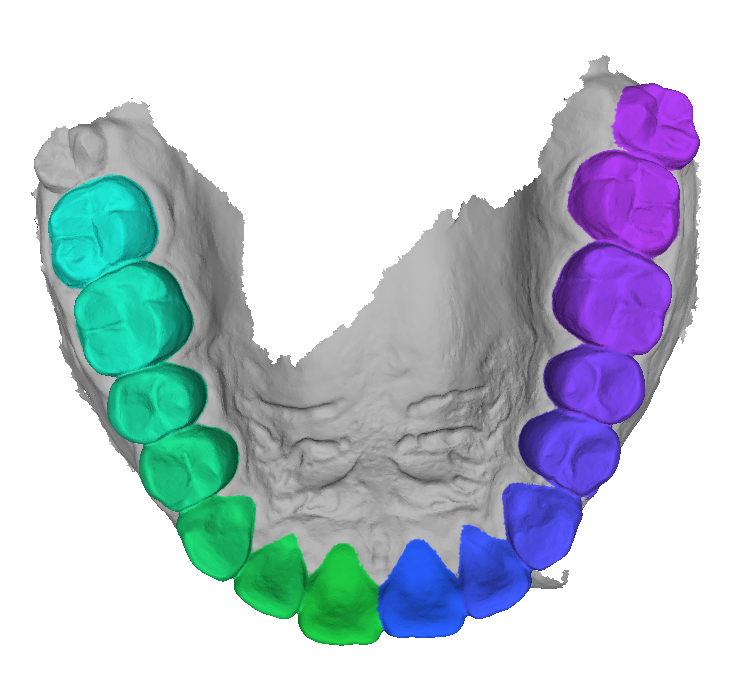}
        \caption{}
        \label{fig:ios}
    \end{subfigure}
    \caption{Illustration of data. (a) Incisors segmented from a CBCT scan, showing complete root anatomy. (b) Full tooth mesh obtained by extracting segmented CBCT surface. (c) IOS mesh, with relatively more precise crown details, but no root information.}
    \label{fig:cbct_ios}
\end{figure}

However, no single imaging modality in clinical use can capture both of these elements adequately. On one hand, CBCT is indispensable for visualizing sub-gingival anatomy, providing essential information about the tooth root and surrounding bone structure (Fig.~\ref{fig:cbct_ios}\subref{fig:cbct}). Yet, CBCT-derived surface models of the tooth crown suffer from significant limitations, including low spatial resolution, noise, and artifacts caused by beam hardening and metallic restorations, rendering them inadequate for applications requiring high precision. On the other hand, IOS generate highly accurate and detailed 3D meshes of the clinical crowns and adjacent soft tissues (Fig.~\ref{fig:cbct_ios}\subref{fig:ios}). Their non-invasive, radiation-free nature has made them the gold standard for digital impressions. The fundamental limitation of IOS, however, is that it is a surface-only technology, incapable of capturing any sub-gingival information, leaving the tooth root completely out of the representation \cite{Mangano2019IOSAccuracy,Alkadi2023IOSAccuracyReview}. This data dichotomy calls for fusion methods that take into account the strengths of both modalities in producing high-fidelity dental models.

\begin{figure}[t]
    \centering
    \begin{subfigure}[b]{0.12\linewidth}
        \centering
        \includegraphics[width=\textwidth]{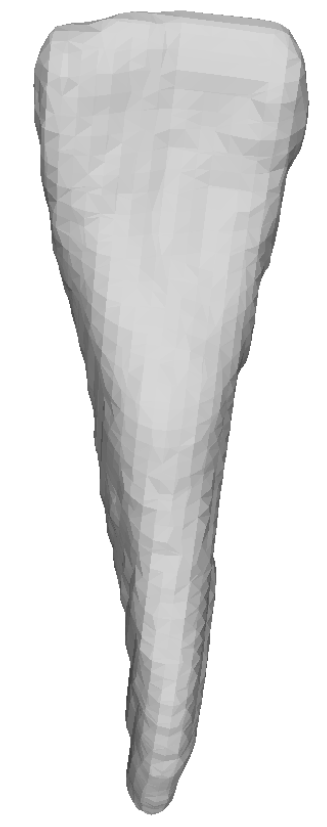}
        \caption{}
        \label{fig:single_cbct}
    \end{subfigure}
    \hfill 
    \begin{subfigure}[b]{0.11\linewidth}
        \centering
        \includegraphics[width=\textwidth]{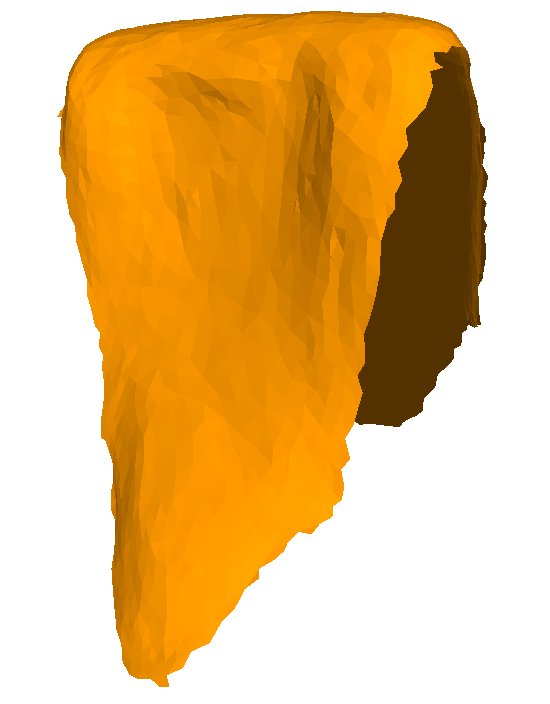}
        \caption{}
        \label{fig:single_ios}
    \end{subfigure}
    \hfill
    \begin{subfigure}[b]{0.12\linewidth}
        \centering
        \includegraphics[width=\textwidth]{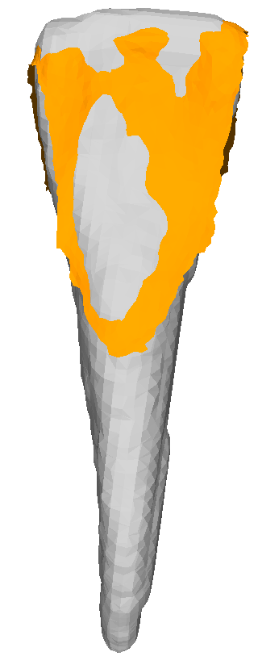}
        \caption{}
        \label{fig:regi}
    \end{subfigure}%
    \hfill 
    \begin{subfigure}[b]{0.12\linewidth}
        \centering
        \includegraphics[width=\textwidth]{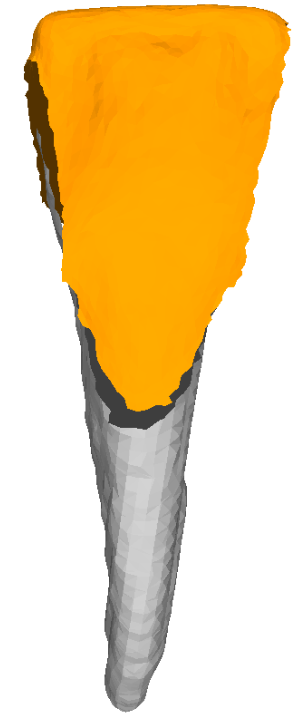}
        \caption{}
        \label{fig:fuse}
    \end{subfigure}
    
    \caption{A naive method for fusing a lower-resolution CBCT root (a, grey) with a high-resolution IOS crown (b, orange) via direct stitching (d). (c) shows the alignment of the two parts. The final model (d) exhibits an unnatural seam, geometric discontinuities, and surface noise at the junction.}
    \label{fig:fusion}
\end{figure}

Prior work has explored tooth segmentation from CBCT using both traditional and deep learning-based methods, with networks such as nnU-Net demonstrating state-of-the-art performance \cite{toothseg,Isensee2021nnUNet}. Surface registration techniques, Random Sample Consensus (RANSAC) and Iterative Closest Point (ICP) algorithm, are commonly used to align different 3D datasets \cite{ransac,Fischler1981RANSAC,Besl1992ICP}. While these tools allow for the creation of a hybrid mesh by simply ``stitching'' the IOS crown to the CBCT root, this naive fusion often results in a non-manifold mesh with an unnatural seam, surface discontinuities, and noise inherited from the CBCT data (Fig.~\ref{fig:fuse}). To address shape completion and regularization, implicit neural representations have emerged as a powerful paradigm. Methods such as DeepSDF~\cite{Park2019DeepSDF} learn a continuous signed distance function (SDF) $f_\theta(\mathbf{x}, \mathbf{z}) = s$ for points in a 3D space $\mathbf{x}\in\mathbb{R}^3$ given a shape latent code $\mathbf{z}\in\mathbb{R}^d$. DeepSDF learns a prior over a class of shapes enabling the reconstruction of high-quality, watertight surfaces from incomplete or noisy data. This surface is given by the zero level set of the learned distance \( \{\mathbf{x}\mid f_\theta(\mathbf{x},\mathbf{z})=0\} \).

In this paper, we propose a novel, fully-automated workflow, as illustrated in Fig.~\ref{fig:pipeline_compact_seg}, for high-fidelity tooth reconstruction that leverages the complementary nature of CBCT and IOS data and harnesses the regularizing power of a deep implicit model. Our contributions are the following:
\begin{itemize}
    \item An automated workflow that integrates SOTA deep learning for tooth instance segmentation from CBCT and IOS.
    \item A robust registration and fusion method to create an initial hybrid mesh of IOS crown with CBCT root.
    \item The core novelty of our work: using a pre-trained DeepSDF network to refine this hybrid mesh, correcting artifacts and generating a single, seamless, and anatomically coherent tooth model that is both complete and highly detailed.
\end{itemize}

\section{Method}
\label{sec:method}

\newdimen\BlockW   \BlockW=.90\linewidth
\newdimen\PairW    \PairW=.48\BlockW
\newdimen\ThumbD   \ThumbD=10.5mm
\newdimen\Gap      \Gap=1.2mm
\newdimen\PairGap  \PairGap=1mm
\newdimen\TextW    \TextW=\dimexpr \PairW-\ThumbD-\Gap\relax
\newdimen\BlockTextW \BlockTextW=\dimexpr \BlockW-\ThumbD-\Gap\relax
\newdimen\SubH     \SubH=11mm
\newdimen\BlockH   \BlockH=11mm
\newdimen\Vone     \Vone=23mm
\newdimen\Vstep    \Vstep=15mm

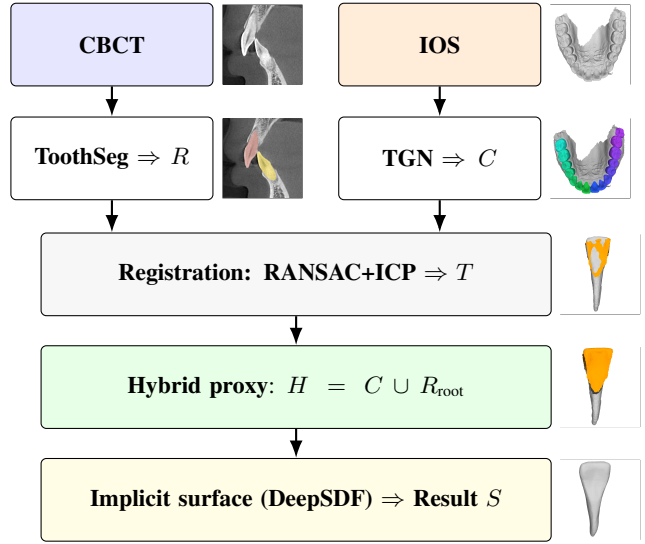
\begin{figure}[t]
\centering
\begin{tikzpicture}[>=Latex]
\usetikzlibrary{matrix,positioning,calc,arrows.meta}

\tikzset{
  sub/.style={draw, rounded corners=2pt, align=center, font=\small,
              inner sep=2pt, minimum height=\SubH, text width=\TextW, fill=white},
  block/.style={draw, rounded corners=2pt, align=center, font=\small,
                inner sep=3pt, minimum height=\BlockH, text width=\BlockTextW, fill=gray!6},
  line/.style={-Latex, thick},
  thumb/.style={draw, outer sep=0pt, minimum width=\ThumbD, minimum height=\ThumbD, inner sep=0pt}
}

\matrix (A) [matrix of nodes, row sep=4mm, column sep=\Gap, nodes={anchor=west}]
{
\node[sub, fill=blue!10] (cbct) {\textbf{CBCT}}; &
\node[thumb] (cbct_t) {\includegraphics[width=\ThumbD,height=\ThumbD]{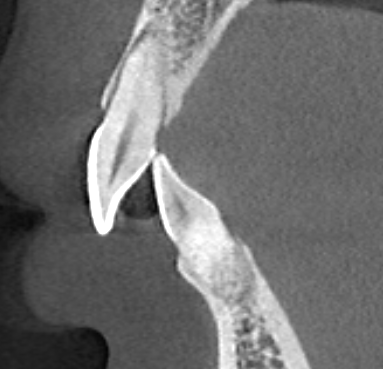}}; &
\node[draw=none, minimum width=\PairGap] {}; &
\node[sub, fill=orange!15] (ios) {\textbf{IOS}}; &
\node[thumb] (ios_t) {\includegraphics[width=\ThumbD,height=\ThumbD]{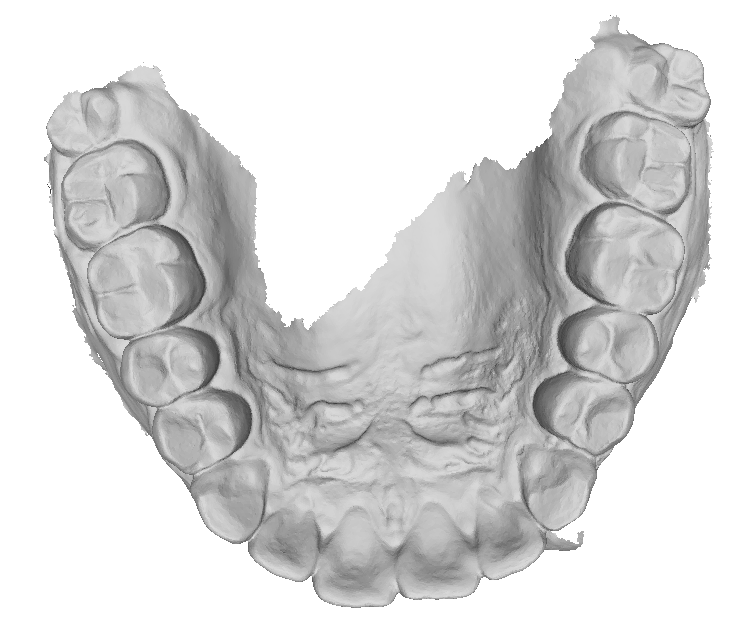}}; \\
\node[sub] (cbctseg) {\textbf{ToothSeg} $\Rightarrow R$}; &
\node[thumb] (cbctseg_t) {\includegraphics[width=\ThumbD,height=\ThumbD]{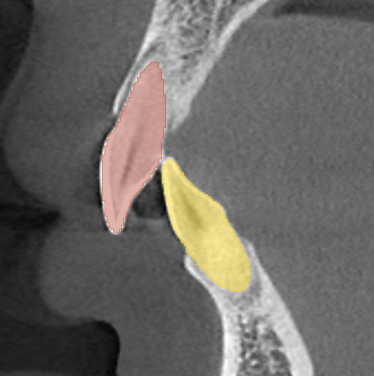}}; &
\node[draw=none, minimum width=\PairGap] {}; &
\node[sub] (iosseg) {\textbf{TGN} $\Rightarrow C$}; &
\node[thumb] (iosseg_t) {\includegraphics[width=\ThumbD,height=\ThumbD]{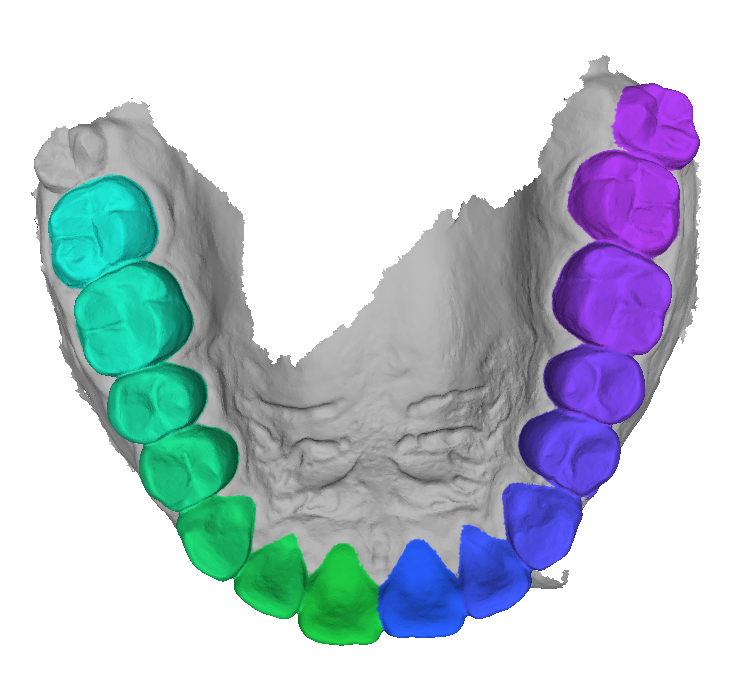}}; \\
};

\draw[line] (cbct.south) -- (cbctseg.north);
\draw[line] (ios.south)  -- (iosseg.north);

\node[thumb, anchor=east] (reg_t)   at ($ (A.east) + (0,-\Vone) $)
  {\includegraphics[width=\ThumbD,height=\ThumbD]{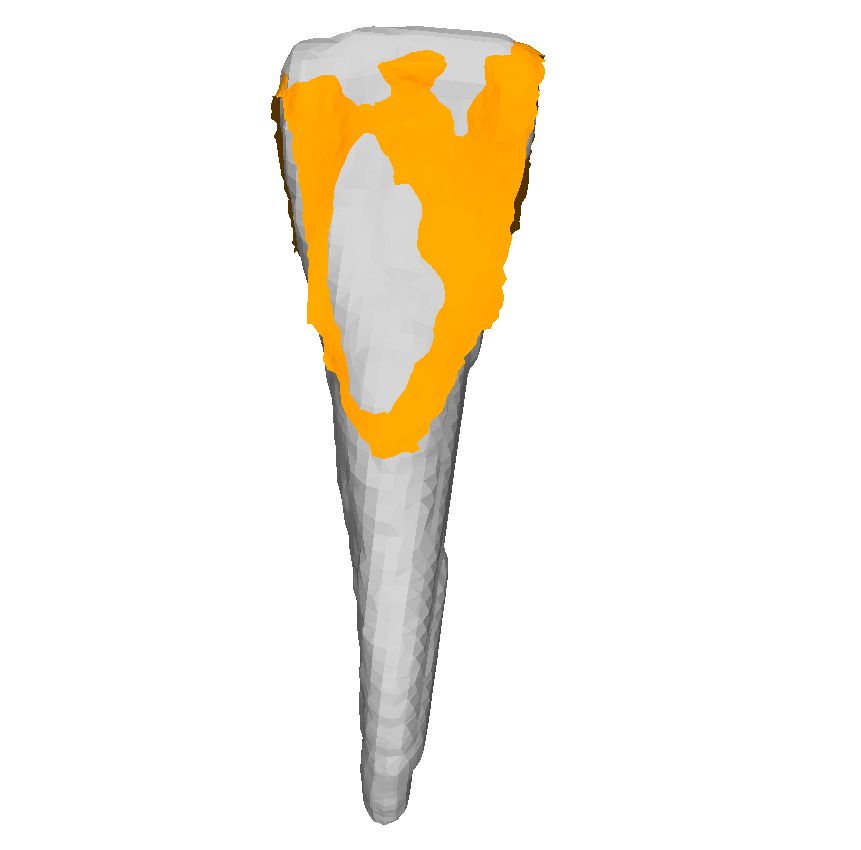}};
\node[block, anchor=east] (reg)     at ($ (A.east) + (-\ThumbD-\Gap,-\Vone) $)
  {\textbf{Registration: RANSAC+ICP} $\Rightarrow$  $T$};
\draw[line] (cbctseg.south) -- ++(0,-2mm) -| (reg.north -| cbctseg);
\draw[line] (iosseg.south)  -- ++(0,-2mm) -| (reg.north -| iosseg);

\node[thumb, anchor=east] (hyb_t)   at ($ (A.east) + (0,-\Vone-\Vstep) $)
  {\includegraphics[width=\ThumbD,height=\ThumbD]{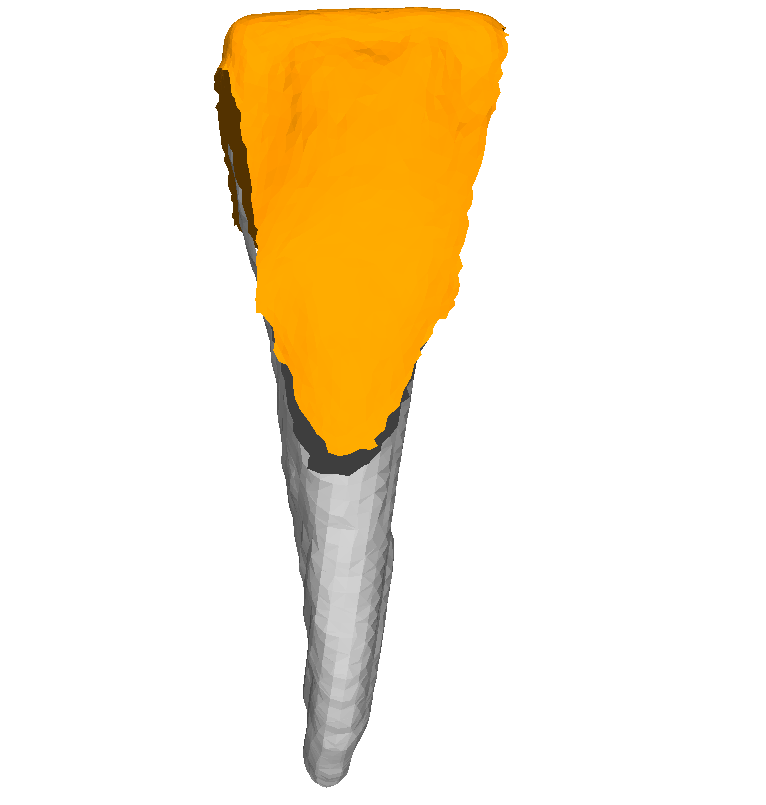}};
\node[block, anchor=east, fill=green!10] (hyb)
  at ($ (A.east) + (-\ThumbD-\Gap,-\Vone-\Vstep) $)
  {\textbf{Hybrid proxy}: $H = C \cup R_{\text{root}}$};
\draw[line] (reg.south) -- (hyb.north);

\node[thumb, anchor=east] (final_t) at ($ (A.east) + (0,-\Vone-2*\Vstep) $)
  {\includegraphics[width=\ThumbD,height=\ThumbD]{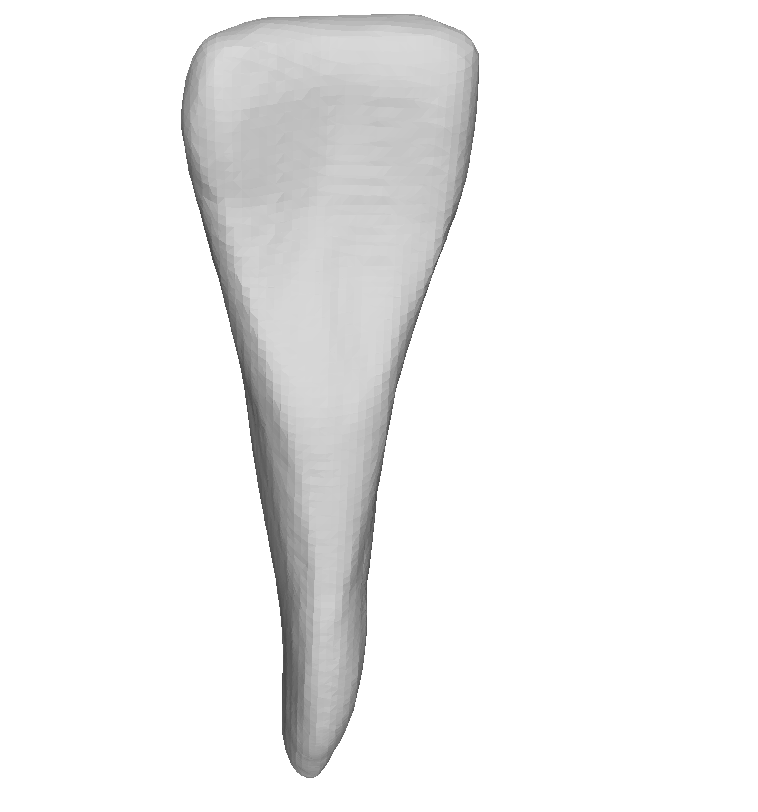}};
\node[block, anchor=east, fill=yellow!12] (final)
  at ($ (A.east) + (-\ThumbD-\Gap,-\Vone-2*\Vstep) $)
  {\textbf{Implicit surface (DeepSDF)} $\Rightarrow$ \textbf{Result} $S$};
\draw[line] (hyb.south) -- (final.north);

\end{tikzpicture}
\caption{Two-column pipeline with an additional segmentation stage. Thumbnails to the right of each box show representative outputs. Abbreviations: $R$ = CBCT segmentation; $C$ = IOS segmentation; $T$ = two-step registration (RANSAC followed by ICP) aligning IOS to CBCT; $H$ = hybrid proxy from naive fusion of crown $C$ with root region $R_{\text{root}}$; $S$ = final implicit surface reconstructed by DeepSDF.}

\label{fig:pipeline_compact_seg}
\end{figure}

\subsection{Data Preparation and Instance Extraction}
\label{ssec:extraction}
We obtain per-tooth instances from both modalities from a publicly available multimodal dataset by Li et al \cite{li2024figshare} by using strong off-the-shelf segmenters. For CBCT, a pre-trained model called ToothSeg~\cite{toothseg} from the nnU-Net family with patch size $256^3$ produces a 3D labelmap of the dentition. We then extract the surface of the target tooth by collecting its 6-neighborhood boundary voxels in physical coordinates and applying a Poisson surface reconstruction with a depth of 10 \cite{Kazhdan2013ScreenedPoisson}. For IOS, a pre-trained mesh labeler named Tooth Group Network~\cite{benhamadou20233dteethseg223dteethscan} isolates the crown triangles of the target tooth. This step yields an IOS crown mesh, denoted as $C$, and a full CBCT tooth mesh, denoted as $R$.

We also extract the surface of the CBCT references from ToothFairy3~\cite{2025CVPR} by the same Poisson surface reconstruction. These surfaces are prepared for future DeepSDF's training.

\subsection{Robust Two-Stage Mesh Registration and Fusion}
\label{ssec:registration}
To accurately align the full CBCT tooth mesh ($R$) to the IOS crown ($C$), we perform a two-stage registration. First, a robust coarse alignment is established by applying a RANSAC-based estimator on Fast Point Feature Histogram (FPFH) features  \cite{FPFH} computed from down-sampled point clouds of the two meshes. This provides a strong initial transformation that is resilient to noise and partial overlap. Second, this alignment is precisely refined using a multi-scale, point-to-plane Iterative Closest Point (ICP) algorithm \cite{Besl1992ICP}, which is responsible for fine-tuning the registration and alignment of the two, yield the final transformation, $T$.

With the meshes aligned, a hybrid mesh is generated to serve as the geometric proxy for the final refinement stage.The root portion of the aligned CBCT mesh is first isolated. This is achieved by classifying each CBCT vertex as belonging to the root if its minimum distance to the IOS crown surface exceeds a threshold of $\tau=0.6$\,mm, which prevents some CBCT residue from extending into the IOS in certain situations. To improve robustness against segmentation noise, we retain only the largest connected component of the resulting root mesh, $R_{\text{root}}$. A naive hybrid mesh ($H$) is then formed by the simple union of the IOS crown ($C$) and the extracted CBCT root ($R_{\text{root}}$). As illustrated in Fig.~\ref{fig:fuse}, this direct stitching results in a mesh with a visible seam, gaps, and other topological artifacts.

\subsection{Implicit Refinement with a Class-Specific DeepSDF}
\label{ssec:deepsdf}

The final stage of our workflow addresses the artifacts present in the hybrid proxy $H$ by leveraging a learned shape prior. We employ a DeepSDF network \cite{Park2019DeepSDF} trained specifically on surfaces extracted from segmentation ground truth from the CBCT dataset ToothFairy3~\cite{2025CVPR}. This network has learned to represent any valid incisor shape as a continuous signed distance function (SDF) controlled by a low-dimensional latent vector $\mathbf{z}$.

During inference, we use the hybrid mesh $H$ as a geometric target. An optimization process is performed to find the latent vector $\mathbf{z}^*$ that causes the implicit function's zero-level set to best fit the surface of $H$. This effectively ``projects'' our noisy, non-manifold input onto the learned space of ideal, complete tooth shapes, correcting the seam, filling holes, and regularizing the surface. Once the optimal $\mathbf{z}^*$ is determined, we extract the final, high-fidelity mesh using the Marching Cubes algorithm on a $192^3$ grid \cite{Lorensen1987MarchingCubes}.

\section{Experiments and Results}
\label{sec:results}

\begin{figure}[t]
  \centering

  \begin{subfigure}[b]{0.14\linewidth}
    \centering
    \includegraphics[width=\textwidth]{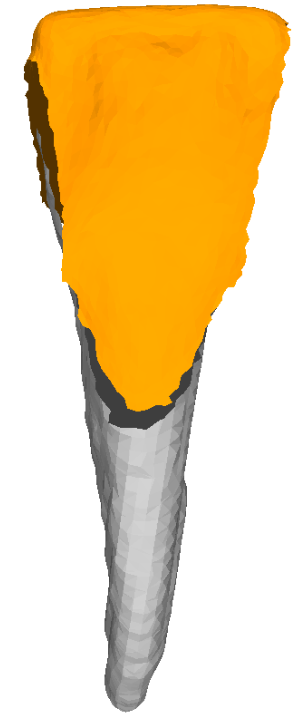}
    \caption{}
    \label{fig:raw}
  \end{subfigure}\hfill
  \begin{subfigure}[b]{0.155\linewidth}
    \centering
    \includegraphics[width=\textwidth]{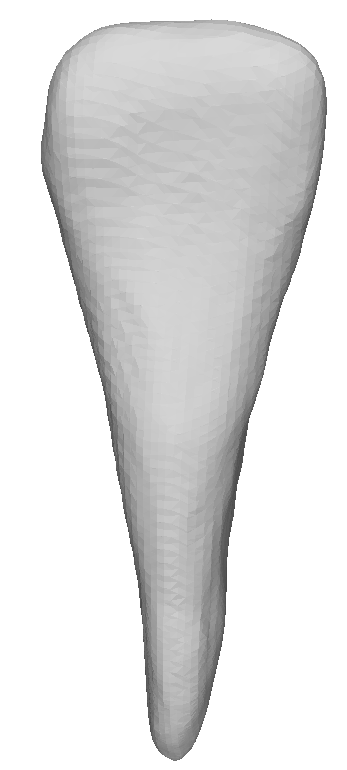}
    \caption{}
    \label{fig:eraw}
  \end{subfigure}\hfill
  \begin{subfigure}[b]{0.145\linewidth}
    \centering
    \includegraphics[width=\textwidth]{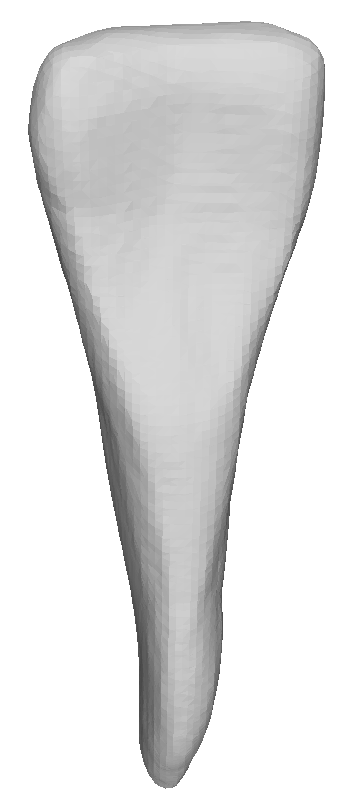}
    \caption{}
    \label{fig:efuse}
  \end{subfigure}\hfill
  \begin{subfigure}[b]{0.13\linewidth}
    \centering
    \includegraphics[width=\textwidth]{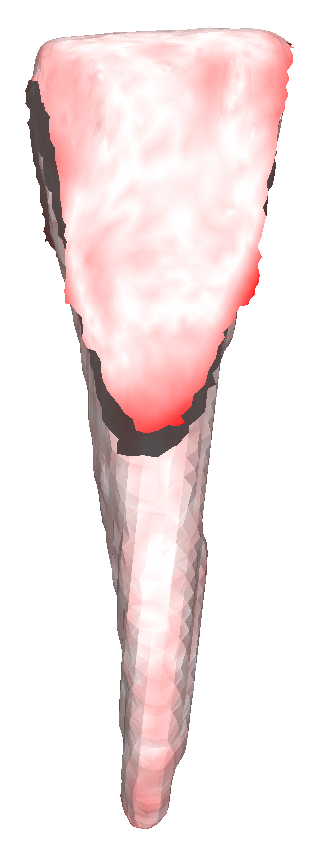}
    \caption{}
    \label{fig:err_fused}
  \end{subfigure}\hfill
  \begin{subfigure}[b]{0.15\linewidth}
    \centering
    \includegraphics[width=\textwidth]{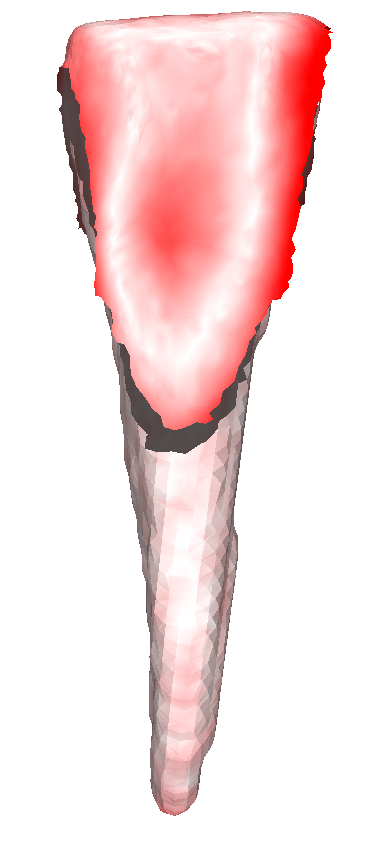}
    \caption{}
    \label{fig:err_cbct}
  \end{subfigure}\hfill
  \begin{subfigure}[b]{0.1\linewidth} 
    \centering
    \includegraphics[width=\textwidth]{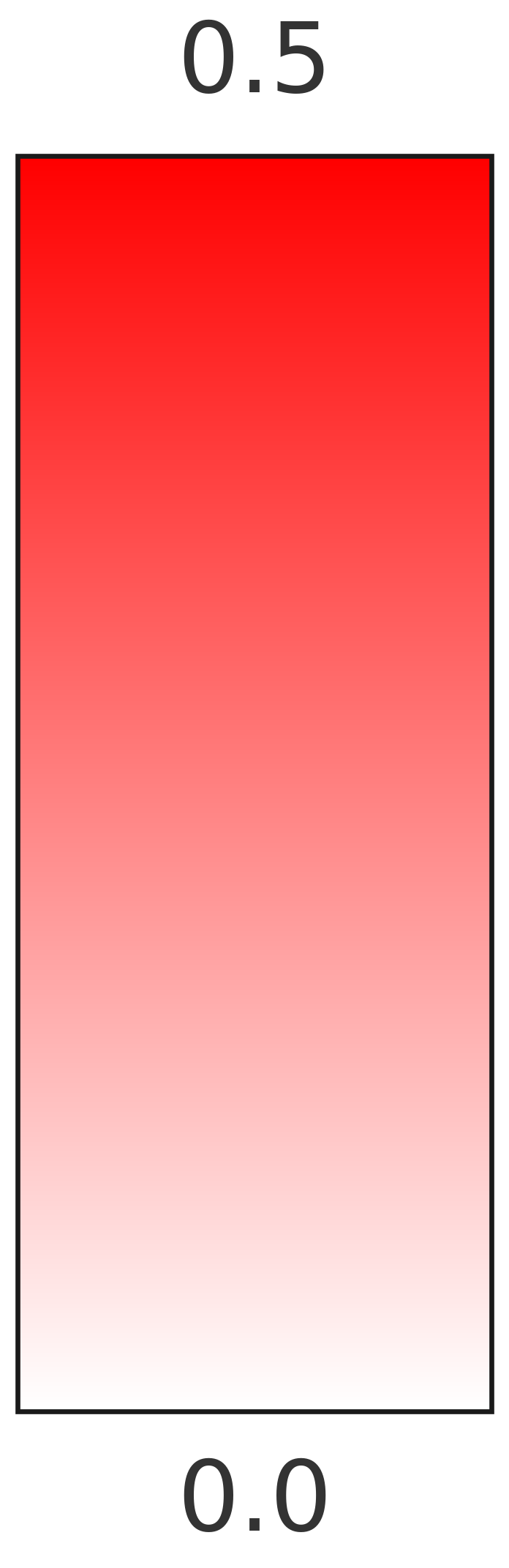}
    \caption{}
    \label{fig:legend}
  \end{subfigure}

  \caption{Qualitative comparison and error maps. Colors represent one-sided surface distance from the Naive-Fusion mesh: \textbf{white = low error}, \textbf{red = high error}.
  (a) Naive-Fusion mesh.
  (b) CBCT-SDF.
  (c) Fused-SDF.
  (d) Error map (CBCT-SDF).
  (e) Error map (Fused-SDF).
  (f) Legend.}
  \label{fig:results}
\end{figure}

\subsection{Qualitative Comparison}
\label{ssec:qualitative}
Fig.~\ref{fig:results} provides a qualitative comparison of the reconstruction methods. In this experiment, the
Naive-Fusion mesh (Fig.~\ref{fig:raw}), which is obtained by rigidly aligning the IOS crown and the CBCT root, serves
as our anatomical reference. Although Naive-Fusion may contain small seams or topological artifacts at the crown--root junction, it is the only available geometry that combines the high-fidelity IOS crown with the CBCT root, we therefore treat it as a practical pseudo-reference.

To visualize reconstruction errors, we compute a one-sided Chamfer distance from
the Naive-Fusion mesh to each DeepSDF reconstruction and map this distance to color. We use a white--to--red
color scheme: white indicates low error (minimal deviation from the reference surface), while increasingly
saturated red indicates high error (larger deviations). We deliberately use a one-sided distance from the Naive-Fusion surface to the reconstructed surfaces in order
to avoid artifacts at the crown--root junction. Naive-Fusion can contain small gaps or seams around the fusion
boundary, which are not meaningful from an anatomical or clinical perspective. By only measuring distances from
the reference surface to the reconstructions, these gaps are not penalized, and the error maps focus
on true geometric discrepancies.

The CBCT-SDF baseline (Fig.~\ref{fig:eraw}) applies DeepSDF regularization directly to the CBCT-based input. In the corresponding error map (Fig.~\ref{fig:err_cbct}), the
crown region exhibits extended red patches, especially in cusp depressions, at the occlusal surface, and along
the lateral walls, indicating substantial deviation from the Naive-Fusion reference crown.

Our Fused-SDF reconstruction (Fig.~\ref{fig:efuse}), which applies DeepSDF regularization to the Naive-Fusion
input, shows a markedly different pattern. The error map in Fig.~\ref{fig:err_fused} is almost entirely white in
the crown, with only small, localized red streaks, indicating that the reconstructed crown closely follows the
high-resolution IOS geometry embedded in the Naive-Fusion reference. The root region also remains predominantly
white for Fused-SDF, confirming that the patient-specific CBCT root morphology is preserved.

Overall, these qualitative observations mirror the quantitative results given in the following section: Fused-SDF achieves substantially lower
error than CBCT-SDF in the crown while maintaining comparable accuracy in the root. This confirms that our
fusion workflow successfully integrates the superior IOS crown with the reliable CBCT root into a coherent,
patient-specific 3D model.

\subsection{Quantitative Analysis}
\label{ssec:quantitative_mandible}

We conducted a quantitative analysis to assess the fidelity of our reconstructions against the Naive-Fusion mesh, which serves as our target reference (IOS crown + CBCT root). For each reconstructed tooth, we compute
one-sided distances from the Naive-Fusion surface to the reconstructed surface, in line with the error maps of
Section~\ref{ssec:qualitative}. Table~\ref{tab:mandible_quant_compare} reports the resulting metrics for the two methods.

\begin{table}[htbp]
\centering
\caption{Quantitative comparison of the Naive-Fusion mesh against the reconstructions. Distances are
one-sided measures from the Naive-Fusion surface to each reconstruction.}
\label{tab:mandible_quant_compare}
\scriptsize
\setlength{\tabcolsep}{5pt}
\begin{tabular}{lccc}
\toprule
\textbf{Input Type} & \textbf{CD(L1)$_{\text{one-side}}$} [mm] & \textbf{HD95$_{\text{one-side,max}}$} [mm] & \textbf{Scale ratio} \\
\midrule
\texttt{CBCT-SDF}  & 0.096$\pm$0.031 & 0.305$\pm$0.067 & 0.983$\pm$0.009 \\
\textbf{\texttt{Fused-SDF} (ours)} & \textbf{0.082$\pm$0.024} & \textbf{0.243$\pm$0.077} & \textbf{0.984$\pm$0.016} \\
\bottomrule
\end{tabular}
\end{table}

The results in Table~\ref{tab:mandible_quant_compare} quantify the reconstruction fidelity against the
Naive-Fusion reference mesh, which represents our desired target geometry (high-fidelity IOS crown combined with
the CBCT root). The CBCT-SDF baseline, which only processes the raw CBCT input without IOS guidance, shows a mean
Chamfer Distance (CD) of 0.096\,mm. This error reflects its deviation from the Naive-Fusion target in
the crown region, which lacks the high-resolution IOS geometry.

In contrast, our Fused-SDF method achieves a lower mean error (CD: 0.082\,mm) and a superior 95th-percentile
Hausdorff distance (HD95: 0.243\,mm), indicating fewer large local deviations. The scale ratio---defined as the
ratio between the bounding-box diagonal of the reconstruction and that of the Naive-Fusion reference---remains
close to unity for both methods (0.983--0.984), confirming that neither approach introduces significant global
scaling artifacts.

Taken together, these quantitative results demonstrate that our fusion network successfully integrates
information from both modalities, producing a final model that is closer to the Naive-Fusion
target than the CBCT-only baseline.

\section{Conclusion}
\label{sec:conclusion}

This paper addresses a critical challenge in digital dentistry: the creation of a single, high-fidelity 3D tooth
model that captures both the detailed crown from IOS scans and the complete root from CBCT data. We have
proposed a novel, fully automated method that leverages a class-specific DeepSDF network to resolve the
artifacts, seams, and topological errors inherent in a naive fusion of these two modalities.

By using the Naive-Fusion mesh (IOS crown + CBCT root) as a clinically meaningful reference, and by designing our
training and evaluation around one-sided surface distances, we have shown that our method produces seamless,
watertight reconstructions that closely follow the desired target anatomy. Our Fused-SDF reconstruction
consistently preserves the patient-specific root morphology while recovering a high-fidelity crown from IOS, as
evidenced by both our qualitative error maps and quantitative metrics.

This approach overcomes the limitations of each imaging modality in isolation and delivers clinically viable 3D
models that are ready for advanced dental applications such as treatment planning, surgical simulation, and
prosthetic design.



\bibliographystyle{IEEEbib}
\bibliography{strings,refs}

\end{document}